\documentclass[a4paper, 11pt]{article}
\usepackage{amsmath,amssymb,color}
\usepackage[dvips]{graphicx}
\usepackage{verbatim}%comment
\usepackage{amsthm}%lemma, proposition, definition

\usepackage{bm}

\theoremstyle{plain}

\newcommand{\R}{\mathbb{R}}
\newcommand{\lambdab}{\pmb\lambda}
\newcommand{\thetab}{\pmb\theta}
\newcommand{\mub}{\pmb\mu}
\newcommand{\vb}{\bf{v}}
\newcommand{\Sb}{\bf S}

\def\heiko{\mathrel{\raise.3ex\hbox{\scalebox{.7}{%
    \rotatebox[origin=c]{-7}{/}%
    \kern-.35em\rotatebox[origin=c]{-7}{/}}}}}%

\title{A tractable, parsimonious and  flexible model for cylindrical data, with applications} 
\author{Toshihiro Abe$^{1}$ and Christophe Ley$^{2}$
\\
\small{$^{1}$ Nanzan University, Nagoya, Japan}
\\
\small{E-mail: abetosh@ss.nanzan-u.ac.jp} 
\\ 
\small{${}^{2}$ Ghent University, Gent, Belgium}
\\ 
\small{E-mail: christophe.ley@ugent.be}
}
\date{}
\begin{document}
\pagenumbering{roman}
\pagenumbering{arabic}
\maketitle

\begin{abstract}
\indent In this paper, we propose cylindrical distributions obtained by combining the sine-skewed von Mises distribution (circular part) with the Weibull distribution (linear part).  This new model, the {WeiSSVM}, enjoys numerous advantages: simple normalizing constant and hence very tractable density, parameter-parsimony and interpretability, good circular-linear dependence structure, easy random number generation thanks to known marginal/conditional distributions,  flexibility illustrated via excellent fitting abilities, and a straightforward extension to the case of directional-linear data. Inferential issues, such as  independence testing, circular-linear respectively linear-circular regression, can easily be tackled with our model, which we apply on two real data sets.   We conclude the paper  by discussing future applications of our model. \end{abstract}


\section{Introduction}\label{sec:Intro}

Cylindrical data are observations that consist of a directional part (a set of angles), which is often of a circular nature (a single angle), and a linear part (mostly a positive real number). This explains the alternative terminology of directional-linear or circular-linear data. Such data occur frequently in natural sciences; typical examples are wind direction and another climatological variable such as wind speed or air temperature, the direction an animal moves and the distance moved, or wave direction and wave height. 
Recent studies of cylindrical data include the exploration of wind direction and SO2 concentration (\cite{GCG13}), the analysis of Japanese earthquakes  (\cite{WSU13}), the link between wildfire orientation and burnt area (\cite{GBCGP14}), and space-time modeling of sea currents in the Adriatic Sea (\cite{WGJ15}, \cite{LPMC15}).
%, and the analysis of tree crown  vectors (\cite{ASAKK15}).

A non-trivial yet fundamental problem is the joint modeling  of the directional/circular and linear variables via the construction of cylindrical probability distributions. The best known examples stem from Mardia and Sutton~(1978) \cite{MS78}, conditioning from a trivariate normal distribution, and Johnson and Wehrly (1978)~\cite{JW78}, invoking maximum entropy principles. The latter also provide in their paper a general way, based on copulas, to construct circular-linear distributions with specified marginals.  

What desirable properties should a ``good'' cylindrical distribution possess? It should  be able to model diverse shapes, in other words present good fitting aptitudes, yet it should ideally remain of a tractable form (this is crucial for  stochastic properties, estimation purposes, and  circular-linear regression) and be parsimonious in terms of parameters at play. The marginal and conditional distributions should optimally be  well-known and flexible  (e.g., there is no reason for the circular component to be always symmetric), whilst the dependence structure has to take care of a reasonable joint behavior. Indeed, numerous examples of cylindrical data require that the circular concentration  tends to increase  with the linear component, as identified in the seminal paper \cite{FL92}. 

All these conditions are well fulfilled by the new model we propose in the present paper. Its  probability density function (pdf) is of the form
\begin{eqnarray}\label{WeiSSVM}
(\theta,x)\mapsto \frac{\alpha\beta^\alpha}{2\pi\cosh(\kappa)}  \left(1+\lambda \sin(\theta-\mu)\right) x^{\alpha-1}\exp\left[-(\beta x)^\alpha\left(1 - \tanh(\kappa)\cos(\theta-\mu)\right)\right], 
\end{eqnarray}
where $(x, \theta) \in [0, \infty) \times [-\pi, \pi)$, $\alpha,\beta>0$, $-\pi\le\mu<\pi$, $\kappa\ge 0$  and 
$-1 \le\lambda\le 1$. The  roles of the distinct parameters will be explained in Section~\ref{model}, as well as the construction underpinning~\eqref{WeiSSVM}.   Stochastic properties such as marginal and conditional distributions, random number generation, moment and correlation calculations are presented in Section~\ref{properties}. We will in particular stress the capacity of our new density to model cylindrical data with length-increasing circular concentration. Maximum likelihood estimation and the ensuing efficient likelihood ratio tests (including tests for circular-linear independence) are discussed in Section~\ref{sec:PE}, as well as circular-linear and linear-circular regression. The excellent modeling capacities of our new model are illustrated by means of two real data sets in Section~\ref{sec:Example}. We conclude the paper by some final comments in Section~\ref{final}, including the straightforward extension of~\eqref{WeiSSVM} to the higher-dimensional directional-linear setting.

\section{A new model for circular-linear data: the WeiSSVM}\label{model}
  
Johnson and Wehrly proposed in~\cite{JW78} a very simple  distribution able to fit cylindrical data where the circular concentration increases with the length of the linear part. Their density reads
\begin{equation}\label{JW} 
(\theta,x)\mapsto\frac{\beta}{2\pi\cosh(\kappa)}  \exp\left[-\beta x\left(1 - \tanh(\kappa)\cos(\theta-\mu)\right)\right], 
\end{equation}
 with  $-\pi\le\mu<\pi$, $\beta>0$ and $\kappa\geq0$. The linear conditional density is the (negative) exponential, while the circular conditional density given $X=x$ is of the form
\begin{equation}\label{vM}
 \theta\mapsto\frac{1}{2\pi I_0(x\beta\tanh(\kappa))} \exp\left[\beta x \tanh(\kappa)\cos(\theta-\mu)\right]
\end{equation}
where $I_0(\kappa)$ is the modified Bessel function of the first kind and order zero. The mapping~\eqref{vM} is the popular von Mises density with location $\mu$ and concentration $\beta x \tanh(\kappa)$, often considered as the circular analogue of the normal distribution.  We attract the reader's attention to the fact that we have  slightly reparameterized   the original Johnson-Wehrly parameterization which would correspond to using $\beta$ and $\kappa_1=\beta\tanh(\kappa)$ instead of $\beta$ and $\kappa$, and hence adding the condition that $\kappa_1<\beta$ in view of $\beta/\cosh(\kappa)=(\beta^2-\kappa_1^2)^{1/2}$. With our parameterization we  avoid this condition, which is an advantage for numerical maximization methods.
 
\begin{figure}%[tbh]
\begin{center}
\begin{tabular}{ccc}
\vspace{0.05in} 
(a) $(\kappa, \lambda)=(0, 0)$ & (b) $(\kappa, \lambda)=(0, 0.5)$ & (c) $(\kappa, \lambda)=(0, 1)$ 
\\
\includegraphics[width=5cm,height=5cm]{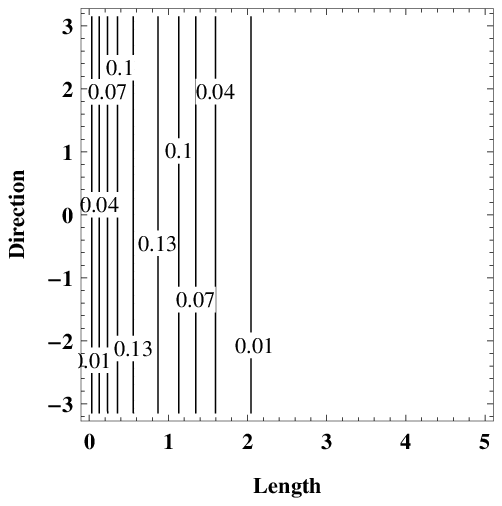}
& 
\includegraphics[width=5cm,height=5cm]{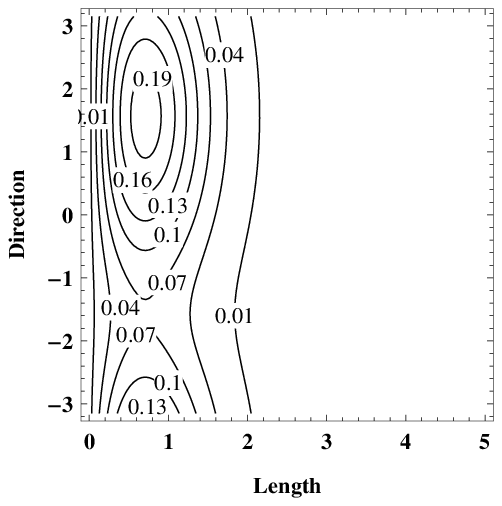}
& 
\includegraphics[width=5cm,height=5cm]{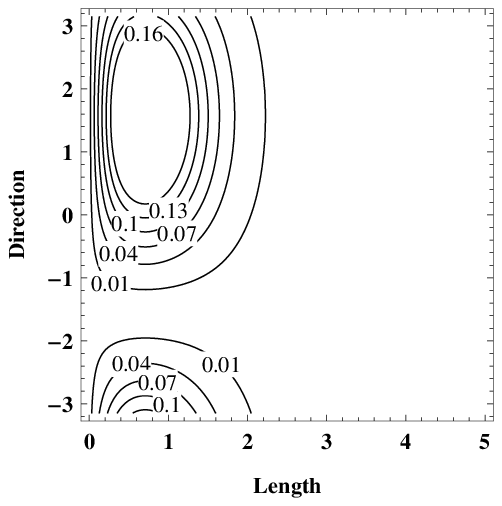}
\\ 
\vspace{0.05in} 
(d) $(\kappa, \lambda)=(1, 0)$ & (e) $(\kappa, \lambda)=(1, 0.5)$ & (f) $(\kappa, \lambda)=(1, 1)$ 
\\
\includegraphics[width=5cm,height=5cm]{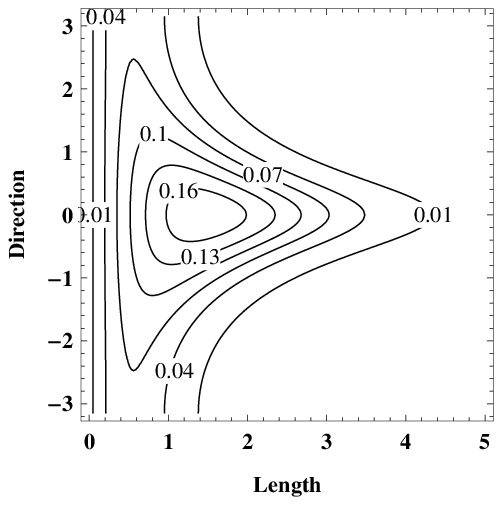}
& 
\includegraphics[width=5cm,height=5cm]{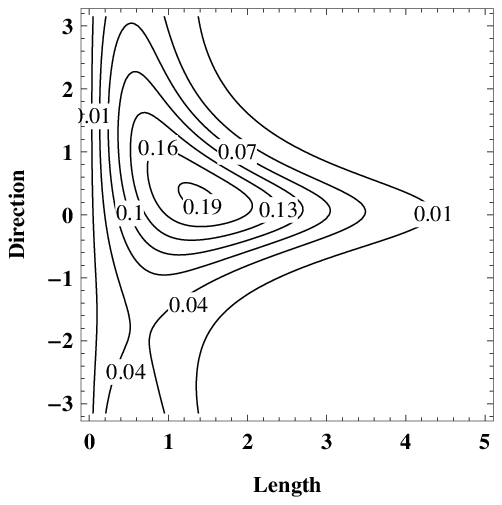}
& 
\includegraphics[width=5cm,height=5cm]{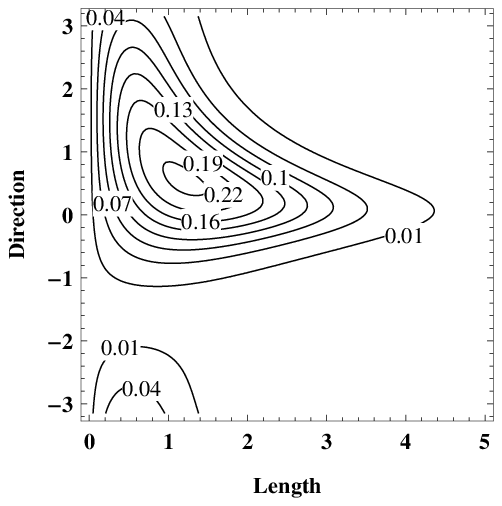}
\\ 
\vspace{0.05in} 
(g) $(\kappa, \lambda)=(1.5, 0)$ & (h) $(\kappa, \lambda)=(1.5, 0.5)$ & (i) $(\kappa, \lambda)=(1.5, 1)$ 
\\
\includegraphics[width=5cm,height=5cm]{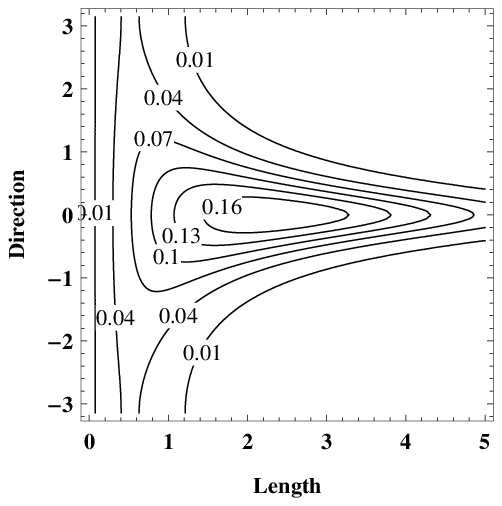}
& 
\includegraphics[width=5cm,height=5cm]{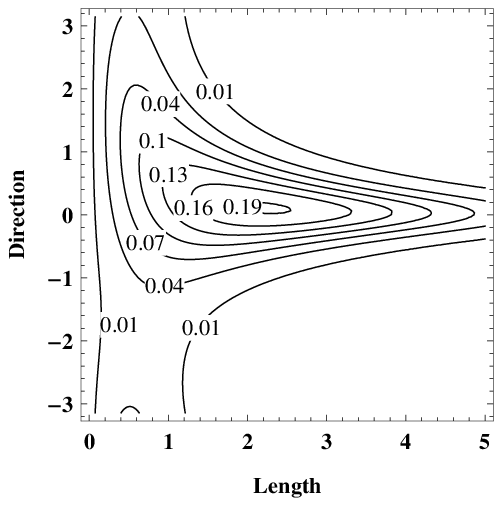}
& 
\includegraphics[width=5cm,height=5cm]{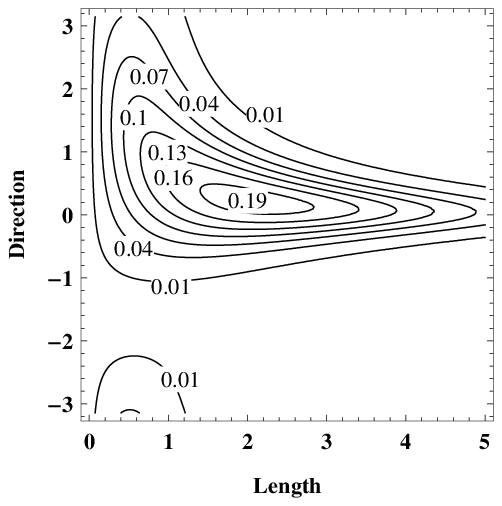}
\end{tabular}
\caption{Contour plots of the {WeiSSVM} density~\eqref{WeiSSVM} over $[0, 5) \times [-\pi, \pi)$ 
for $(\alpha, \beta, \mu)=(2, 1, 0)$ 
with 
(a) $(\kappa, \lambda)=(0, 0)$ (Weibull and uniform) 
(b) $(\kappa, \lambda)=(0, 0.5)$ (Weibull and cardioid), 
(c) $(\kappa, \lambda)=(0, 1)$ (Weibull and cardioid), 
(d) $(\kappa, \lambda)=(1, 0)$, (e) $(\kappa, \lambda)=(1, 0.5)$, (f) $(\kappa, \lambda)=(1, 1)$, 
(g) $(\kappa, \lambda)=(1.5, 0)$, (h) $(\kappa, \lambda)=(1.5, 0.5)$ and (i) $(\kappa, \lambda)=(1.5, 1)$. 
}
\label{fig:Circular-Linear}
\end{center}
\end{figure}

A drawback of the Johnson-Wehrly model~\eqref{JW} is its lack of flexibility. Both its conditional and marginal circular densities are symmetric (see Section~\ref{margcond} for details), when $\kappa=0$ the circular contribution in~\eqref{JW} boils down to the uniform law on $[-\pi,\pi)$, and the circular concentration    varies linearly with $x$ (see~\eqref{vM}). In order to overcome these limitations, we have applied two separate transformations to the Johnson-Wehrly density: a power transformation $x\mapsto x^{1/\alpha}$ for $\alpha>0$ to the linear part, and a perturbation of the circular part via multiplication with $\theta\mapsto (1+\lambda \sin(\theta-\mu))$ for $\lambda\in[-1,1]$. The former is the classical way to turn an exponential distribution on $\R^+$ to the Weibull distribution with  pdf $x\mapsto\alpha \beta x^{\alpha-1}\exp\left[-\beta x^\alpha\right]$, which is a very popular distribution to model diverse natural phenomena. The effect of the  perturbation is  known in circular statistics as ``sine-skewing'' a reflectively symmetric distribution, see \cite{AP11}. Whenever $\lambda\neq0$, the resulting density becomes skewed, whereas symmetry is retrieved for $\lambda=0$; moreover, the perturbation leaves the normalizing constant untouched. The combined effect of both transformations (plus the change from $\beta$ to $\beta^\alpha$ mainly for aesthetic reasons) thus yields the pdf  
$$
\frac{\alpha\beta^\alpha}{2\pi\cosh(\kappa)}  \left(1+\lambda \sin(\theta-\mu)\right) x^{\alpha-1}\exp\left[-(\beta x)^\alpha\left(1 - \tanh(\kappa)\cos(\theta-\mu)\right)\right],
$$
which we term \emph{WeiSSVM} for the interplay between the linear Weibull part and the circular sine-skewed von Mises part. 2D contour plots of the density \eqref{WeiSSVM} are given in Figure \ref{fig:Circular-Linear} and show the versatility of our new model. 

Parameter interpretation becomes now clear: $\mu$ and $\lambda$ respectively endorse the role of circular location and skewness parameters, while $\beta$ and $\alpha$ are linear scale and shape parameters. The parameter $\kappa$ plays, as in the original Johnson-Wehrly model, the role of circular concentration and circular-linear dependence parameter. Independence is attained when $\kappa=0$, in which case the density \eqref{WeiSSVM} becomes the product of the linear Weibull and the circular cardioid distribution with location $\mu+\pi/2$ and concentration $\lambda$, see the first row of Figure \ref{fig:Circular-Linear}.

\section{Properties of the WeiSSVM}\label{properties}

\subsection{The normalizing constant}

As can be seen from~\eqref{WeiSSVM}, the normalizing constant is  very simple, which is a  strong asset of our proposal. Indeed, it is not rare to encounter normalizing constants defined in terms of infinite series of functions, as for the Kato-Shimizu model \cite{KS08}, see also~\eqref{pdf:KS}. 

%Although the normalizing constant is immediately obtained from the transformations we apply to~\eqref{JW}, we  briefly establish its expression:
%\begin{eqnarray*}
%&&
%\int_0^\infty\int_{-\pi}^\pi \left(1+\lambda \sin(\theta-\mu)\right) x^{\alpha-1}\exp\left[-(\beta x)^\alpha\left(1 - \tanh(\kappa)\cos(\theta-\mu)\right)\right]d\theta dx
%\\
%&=&
%\int_0^\infty\int_{-\pi}^\pi x^{\alpha-1}
%\exp\left[-(\beta x)^\alpha\left(1 - \tanh(\kappa)\cos(\theta-\mu)\right)\right]d\theta dx
%\\
%&=&
%\frac{1}{\alpha\beta^\alpha} \int_{-\pi}^\pi\frac{1}{1 - \tanh(\kappa)\cos(\theta-\mu)}d\theta
%\\
%&=&
%\frac{1}{\alpha\beta^\alpha} 
%\int_{-\pi}^\pi
%\frac{1+\tanh^2(\kappa/2)}{1+\tanh^2(\kappa/2) - 2\tanh(\kappa/2)\cos(\theta-\mu)}
%d\theta
%\\
%&=&
%\frac{2\pi\cosh(\kappa)}{\alpha\beta^\alpha},
%\end{eqnarray*}
%where we have used the facts that 
%$(1-\tanh^2(\kappa/2))/[2\pi(1+\tanh^2(\kappa/2) - 2\tanh(\kappa/2)\cos(\theta-\mu))]$ 
%is the density of the wrapped Cauchy distribution 
%and 
%$(1+\tanh^2(\kappa/2))/(1-\tanh^2(\kappa/2))=\cosh(\kappa)$. 

\subsection{Marginal and conditional distributions}\label{margcond}

The  marginal density of the circular component $\Theta$ from pdf (\ref{WeiSSVM})  is 
given by 
\begin{eqnarray*}
f(\theta) 
&=& \frac{1}{2\pi\cosh(\kappa)} \left(1+\lambda \sin(\theta-\mu)\right) 
\int_{0}^{\infty}  \alpha\beta^\alpha x^{\alpha-1}\exp\left[-(\beta x)^\alpha\left(1 - \tanh(\kappa)\cos(\theta-\mu)\right)\right]dx
\\
&=& \frac{1}{2\pi\cosh(\kappa)} \frac{1+\lambda \sin(\theta-\mu)}{1 - \tanh(\kappa)\cos(\theta-\mu)}\\
&=&\frac{1-\tanh^2(\kappa/2)}{2\pi}\frac{1+\lambda \sin(\theta-\mu)}{1+\tanh^2(\kappa/2) - 2\tanh(\kappa/2)\cos(\theta-\mu)},
\end{eqnarray*}
which is the sine-skewed wrapped Cauchy distribution (\cite{AP11}), a  flexible extension of the symmetric wrapped Cauchy distribution.

The  marginal density of the linear component $X$ from pdf (\ref{WeiSSVM})  in turn corresponds to 
\begin{eqnarray*}
f(x) 
&=& 
\frac{1}{2\pi\cosh(\kappa)}\alpha\beta^\alpha x^{\alpha-1} 
\int_{-\pi}^{\pi} \left(1+\lambda \sin(\theta-\mu)\right)  \exp\left[-(\beta x)^\alpha\left(1 - \tanh(\kappa)\cos(\theta-\mu)\right)\right]d\theta\\
&=& \frac{1}{2\pi\cosh(\kappa)}\alpha\beta^\alpha x^{\alpha-1}\exp[-(\beta x)^\alpha] \int_{-\pi}^{\pi}   \exp\left[(\beta x)^\alpha \tanh(\kappa)\cos(\theta-\mu)\right]d\theta\\
&=&\frac{I_0(x^\alpha\beta^\alpha\tanh(\kappa))}{\cosh(\kappa)}\alpha\beta^\alpha x^{\alpha-1}\exp[-(\beta x)^\alpha].
\end{eqnarray*}
This is an extended version of the marginal density of~\eqref{JW} given in \cite{JW78}; 
as already noticed, it simplifies to the Weibull {when} $\kappa=0$.

The conditional densities from (\ref{WeiSSVM}) are now readily given by
\begin{equation}\label{thetax}
f(\theta|x) = 
\frac{1}{2\pi I_0(x^\alpha\beta^\alpha\tanh(\kappa))}\left(1+\lambda \sin(\theta-\mu)\right) \exp\left[(\beta x)^\alpha \tanh(\kappa)\cos(\theta-\mu)\right]
\end{equation}
and
\begin{equation}\label{xtheta}
f(x|\theta) = 
\alpha\left[\beta (1 - \tanh(\kappa)\cos(\theta-\mu))^{1/\alpha}\right]^\alpha x^{\alpha-1}
\exp\left[-\left(\beta \left(1 - \tanh(\kappa)\cos(\theta-\mu)\right)^{1/\alpha}x\right)^\alpha\right].
\end{equation}
Both densities are quite common; \eqref{thetax} is the sine-skewed von Mises distribution with concentration $(\beta x)^\alpha \tanh(\kappa)$ (note how the concentration now varies with $x^\alpha$ instead of simply $x$) whereas \eqref{xtheta} is the Weibull with shape parameter $\beta \left(1 - \tanh(\kappa)\cos(\theta-\mu)\right)^{1/\alpha}$. In general the conditional density \eqref{thetax} is unimodal, but it can become bimodal when the absolute value of the skewness parameter $\lambda$ is large (from our experiment, $|\lambda|>0.918$) 
because of the structure of the sine-skewed von Mises distribution, see~\cite{AP11}.

\subsection{Random number generation}

Thanks to the results of the previous section, we can describe a simple random number generation algorithm by decomposing $f(\theta,x)$ into $f(x|\theta) f(\theta)$, in other words, by first generating $\Theta\sim f(\theta)$ and then $X|\Theta=\theta \sim f(x|\theta)$. The algorithm goes as follows.

\begin{itemize}
\item[Step 1:] Generate a random variable $\Theta_1$ following a (symmetric) wrapped Cauchy law with location $\mu$ and concentration $\tanh(\kappa/2)$, and generate independently $U\sim Unif[0,1]$.
\item[Step 2:] Define $\Theta$ as
$$
\left\{\begin{array}{ll}
\Theta_1& \mbox{if}\,\, U< (1+\lambda \sin(\Theta_1-\mu))/2\\
-\Theta_1& \mbox{if}\,\, U\geq (1+\lambda \sin(\Theta_1-\mu))/2;
\end{array}
\right.
$$ $\Theta$ then follows the sine-skewed wrapped Cauchy distribution.
\item[Step 3:] Generate $X$ from a Weibull with  shape parameter 
$\beta \left(1 - \tanh(\kappa)\cos(\Theta-\mu)\right)^{1/\alpha}$.
\end{itemize}
Random number generation from sine-skewed distributions follows from general skew-symmetric theory on $\R^k$; see \cite{WBG04}.

\subsection{Moment expressions}\label{sec:moments}

The  moments of the Weibull distribution
and trigonometric moments of the sine-skewed von Mises distribution are given  explicitly. 
These nice properties are inherited to our model. 

For $n = 1, 2, \ldots$ and $m = 1, 2, \ldots$, we have 
\begin{eqnarray*} 
&&{\rm E}[X^n \cos(m\Theta)] \\
&=& 
\frac{\alpha\beta^\alpha}{2\pi\cosh(\kappa)} \int_0^\infty\int_{-\pi}^\pi x^n \cos(m\theta)\left(1+\lambda \sin(\theta)\right) x^{\alpha-1}\exp\left[-(\beta x)^\alpha\left(1 - \tanh(\kappa)\cos(\theta)\right)\right] d\theta dx\\ 
&=&\frac{1}{2\pi\cosh(\kappa)} \int_{-\pi}^\pi \cos(m\theta) \int_0^\infty \alpha\beta^\alpha x^n  x^{\alpha-1}\exp\left[-(\beta x)^\alpha\left(1 - \tanh(\kappa)\cos(\theta)\right)\right] dxd\theta 
\\
&=&
\frac{1}{2\pi\cosh(\kappa)} \int_{-\pi}^\pi \cos(m\theta) \frac{\Gamma(1+n/\alpha)}{\beta^{n}\left(1 - \tanh(\kappa)\cos(\theta)\right)^{n/\alpha+1}}d\theta
\\
&=&
\frac{\Gamma(n/\alpha+1)(\cosh(\kappa))^{n/\alpha+1}}{\cosh(\kappa)\beta^{n}}
\int_{-\pi}^\pi \frac{1}{2\pi}\cos(m\theta)\frac{1}{\left(\cosh(\kappa) - \sinh(\kappa)\cos(\theta)\right)^{n/\alpha+1}}d\theta
\\
&=&
\frac{\Gamma(n/\alpha+1)(\cosh(\kappa))^{n/\alpha}}{\beta^{n}}
\frac{\Gamma(n/\alpha+1-m)P^m_{n/\alpha}(\cosh(\kappa))}{\Gamma(n/\alpha+1)}
\\
&=&
\frac{(\cosh(\kappa))^{n/\alpha}\Gamma(n/\alpha+1-m)}{\beta^{n}}
P^m_{n/\alpha}(\cosh(\kappa)),
\end{eqnarray*}
where $P^m_{\nu}(z)$ is the associated Legendre function of 
the first kind of degree $\nu$ and order $m$ given by
(equation 8.711.2 of \cite{GR15}, p. 969) 
$$ 
P^m_{\nu}(z) = 
\frac{(-\nu)_m}{\pi} 
\int^{\pi}_{0} 
\frac{\cos mt}{(z+\sqrt{z^2-1}\cos t)^{\nu+1}} 
dt
= 
\frac{\Gamma(\nu+1)}{\pi\Gamma(\nu-m+1)}
\int^{0}_{-\pi} 
\frac{\cos mt}{(z-\sqrt{z^2-1}\cos t)^{\nu+1}} 
dt. 
$$ 
Here, we used the relation 
$$ 
(-\nu)_m 
= 
\frac{\Gamma(m-\nu)}{\Gamma(-\nu)} 
= 
(-1)^m\frac{\Gamma(\nu+1)}{\Gamma(\nu-m+1)}. 
$$

%where $P_\gamma(\cdot)$ is the Legendre function of the first kind of degree $\gamma$ and ordre 0 (see Gradshteyn and Ryzhik~1994, secs. 8.7 and 8.8). In this calculation we have made use of both the moment expressions of the Weibull distribution and of the formula in Section 2.6 of Jones and Pewsey~(2005), as it turns out that here, as well as in other manipulations, the circular integrant remaining after integrating out over $(0,\infty)$ conveniently reduces to a Jones-Pewsey density.

Similarly,
\begin{eqnarray*}
&&{\rm E}[X^n \sin(m\Theta)] \\
&=& \frac{\alpha\beta^\alpha}{2\pi\cosh(\kappa)} 
\int_0^\infty\int_{-\pi}^\pi x^n \sin(m\theta)\left(1+\lambda \sin(\theta)\right) x^{\alpha-1}
\exp\left[-(\beta x)^\alpha\left(1 - \tanh(\kappa)\cos(\theta)\right)\right] d\theta dx
\\ 
&=&
\frac{\lambda}{2\pi\cosh(\kappa)} \int_{-\pi}^\pi \sin(m\theta)\sin(\theta) \int_0^\infty \alpha\beta^\alpha x^n  x^{\alpha-1}\exp\left[-(\beta x)^\alpha\left(1 - \tanh(\kappa)\cos(\theta)\right)\right] dxd\theta 
\\
&=& 
\frac{\lambda}{2\pi\cosh(\kappa)} 
\int_{-\pi}^\pi \sin(m\theta)\sin(\theta) 
\frac{\Gamma(n/\alpha+1)}{\beta^{n}\left(1 - \tanh(\kappa)\cos(\theta)\right)^{n/\alpha+1}}d\theta
\\
&=&
\frac{\lambda\Gamma(n/\alpha+1)(\cosh(\kappa))^{(n/\alpha+1)}}{\cosh(\kappa)\beta^{n}}
\int_{-\pi}^\pi \frac{1}{2\pi}\frac{(\cos((m-1)\theta)-\cos((m+1)\theta))}{2\left(\cosh(\kappa) - \sinh(\kappa)\cos(\theta)\right)^{n/\alpha+1}}d\theta
\\
&=&
\frac{\lambda\Gamma(n/\alpha+1)(\cosh(\kappa))^{n/\alpha}}{\beta^{n}}
\frac{1}{2}
\left(
\frac{\Gamma(n/\alpha+2-m)}{\Gamma(n/\alpha+1)}
P^{m-1}_{n/\alpha}(\cosh(\kappa))
- 
\frac{\Gamma(n/\alpha-m)}{\Gamma(n/\alpha+1)}
P^{m+1}_{n/\alpha}(\cosh(\kappa))
\right)
\\
&=&
\frac{\lambda(\cosh(\kappa))^{n/\alpha}}{\beta^{n}}
\frac{\left(\Gamma(n/\alpha+2-m)
P^{m-1}_{n/\alpha}(\cosh(\kappa))
- 
\Gamma(n/\alpha-m)
P^{m+1}_{n/\alpha}(\cosh(\kappa))\right)}
{2}. 
\end{eqnarray*}

%The expression for the associated Legendre function in some packages 
%is slightly different from the one given in Gradshteyn and Ryzhik (2015). 
%For example, its expression in {\it Mathematica} corresponds to  
%$$ 
%P^m_{\nu}(z) = 
%\frac{(-\nu)_m}{\pi} e^{\frac{m\pi i}{2}}
%\int^{\pi}_{0} 
%\frac{\cos mt}{(z+\sqrt{z^2-1}\cos t)^{\nu+1}} 
%dt 
%$$ 
%for $m\in\N$ and $\Re z>0$, where $i = \sqrt{-1}$, hence one has to be careful about those differences in the definition when using the package.
%Therefore, if one uses the package, attention has to paid on  whether multiplication  by $i^m$, or not, . 

Specifying choices for $m$ and $n$, and noting that the marginal of the circular part is 
the  sine-skewed wrapped Cauchy density, we obtain  the following simple moment expressions (we write $P_\nu(z)$ for $P^0_{\nu}(z)$)
\begin{eqnarray*}
%%%
{\rm E}[X] 
&=&
\frac{(\cosh(\kappa))^{1/\alpha}\Gamma\left(\frac{1}{\alpha}+1\right)}{\beta}
P_{1/\alpha}(\cosh(\kappa)), 
\\ 
%%%
{\rm E}[X^2] 
&=& 
\frac{(\cosh(\kappa))^{2/\alpha}\Gamma\left(\frac{2}{\alpha}+1\right)}{\beta^2}
P_{2/\alpha}(\cosh(\kappa)),
\\
%%%
{\rm E}[\cos(\Theta)] 
&=&
\tanh\left(\frac{\kappa}{2}\right), 
%%%
\quad 
{\rm E}[\cos^2(\Theta)] 
=
\frac{1}{2}\left(1+\tanh^2\left(\frac{\kappa}{2}\right)\right), 
\\
%%%
{\rm E}[\sin(\Theta)] 
&=&
\frac{\lambda}{2\cosh^2\left(\frac{\kappa}{2}\right)}, 
\quad 
%%%
{\rm E}[\sin^2(\Theta)] 
=
\frac{1}{2\cosh^2\left(\frac{\kappa}{2}\right)}, 
\\
%%%
{\rm E}[X \cos(\Theta)] 
&=&
\frac{(\cosh(\kappa))^{1/\alpha}\Gamma\left(\frac{1}{\alpha}\right)}{\beta}
P^1_{1/\alpha}(\cosh(\kappa)),
\\
%%%
{\rm E}[X \sin(\Theta)] 
&=& 
\frac{\lambda(\cosh(\kappa))^{1/\alpha}}{\beta}
\frac{\left(\Gamma\left(\frac{1}{\alpha}+1\right)
P_{1/\alpha}(\cosh(\kappa))
- 
\Gamma\left(\frac{1}{\alpha}-1\right)
P^{2}_{1/\alpha}(\cosh(\kappa))\right)}
{2}, 
%%%
\\
{\rm E}[\cos(\Theta)\sin(\Theta)] 
&=& 
\frac{\lambda\tanh\left(\frac{\kappa}{2}\right)}{4\cosh^2\left(\frac{\kappa}{2}\right)}. 
\end{eqnarray*}

\subsection{Circular-linear correlation}

From the moment expressions of the previous section we readily derive the following quantities:
\begin{eqnarray*}
%%%%%%
{\rm Var}[X] 
&=&
\frac{(\cosh(\kappa))^{2/\alpha}\left(
\Gamma\left(\frac{2}{\alpha}+1\right)P_{2/\alpha}(\cosh(\kappa)) 
- 
\Gamma\left(\frac{1}{\alpha}+1\right)^2 P_{1/\alpha}(\cosh(\kappa))^2 
\right)}{\beta^2}, 
\\
%%%%%%
{\rm Var}[\cos(\Theta)] 
&=&
\frac{1}{2\cosh^2\left(\frac{\kappa}{2}\right)}, 
\\
%%%%%%
{\rm Var}[\sin(\Theta)] 
&=&
\frac{1}{2\cosh^2\left(\frac{\kappa}{2}\right)} 
\left(
1-\frac{\lambda^2}{2\cosh^2\left(\frac{\kappa}{2}\right)}
\right), 
\\ 
%%%
{\rm Cov}(X, \cos(\Theta))
&=& 
\frac{(\cosh(\kappa))^{1/\alpha}\left(
\Gamma\left(\frac{1}{\alpha}\right)P^1_{1/\alpha}(\cosh(\kappa)) 
- 
\Gamma\left(\frac{1}{\alpha}+1\right)
P_{1/\alpha}(\cosh(\kappa))\tanh\left(\frac{\kappa}{2}\right)
\right)}{\beta}, 
\\ 
%%%
{\rm Cov}(X, \sin(\Theta)) 
&=&
\frac{\lambda(\cosh(\kappa))^{1/\alpha}\left(
\Gamma\left(\frac{1}{\alpha}+1\right)\tanh^2\left(\frac{\kappa}{2}\right)P_{1/\alpha}(\cosh(\kappa))
- 
\Gamma\left(\frac{1}{\alpha}-1\right) P^{2}_{1/\alpha}(\cosh(\kappa))
\right)}{2\beta}, 
\\ 
%%%
{\rm Cov}(\cos(\Theta), \sin(\Theta)) 
&=& 
-\frac{\lambda\tanh\left(\frac{\kappa}{2}\right)}{4\cosh^2\left(\frac{\kappa}{2}\right)}. 
\end{eqnarray*}

Using these expressions, the correlations are given by 
\begin{eqnarray*}
%%%%%%
r_{xc} &=& {\rm Corr}(X, \cos(\Theta))
\\ 
&=& 
\frac{
\sqrt{2}\cosh\left(\frac{\kappa}{2}\right)
\left(
\Gamma\left(\frac{1}{\alpha}\right)P^1_{1/\alpha}(\cosh(\kappa)) 
- 
\Gamma\left(\frac{1}{\alpha}+1\right)
P_{1/\alpha}(\cosh(\kappa))\tanh\left(\frac{\kappa}{2}\right)
\right)}
{
\sqrt{
\Gamma\left(\frac{2}{\alpha}+1\right)P_{2/\alpha}(\cosh(\kappa)) 
- 
\Gamma\left(\frac{1}{\alpha}+1\right)^2 P_{1/\alpha}(\cosh(\kappa))^2 
}
}, 
\\ 
%%%%%%
r_{xs} &=& {\rm Corr}(X, \sin(\Theta))
\\ 
&=& 
\frac{\lambda
\cosh\left(\frac{\kappa}{2}\right)
\left(
\Gamma\left(\frac{1}{\alpha}+1\right)\tanh^2\left(\frac{\kappa}{2}\right)P_{1/\alpha}(\cosh(\kappa))
- 
\Gamma\left(\frac{1}{\alpha}-1\right) P^{2}_{1/\alpha}(\cosh(\kappa))
\right)
}
{
\sqrt{2}\sqrt{
\Gamma\left(\frac{2}{\alpha}+1\right)P_{2/\alpha}(\cosh(\kappa)) 
- 
\Gamma\left(\frac{1}{\alpha}+1\right)^2 P_{1/\alpha}(\cosh(\kappa))^2 
}
\sqrt{
1-\frac{\lambda^2}{2\cosh^2\left(\frac{\kappa}{2}\right)} 
}
}, 
\\ 
%%%%%%
r_{cs} 
&=& 
{\rm Corr}(\cos(\Theta), \sin(\Theta))
\\ 
&=& 
-\frac{
\lambda\tanh\left(\frac{\kappa}{2}\right)
}
{
2\sqrt{
1-\frac{\lambda^2}{2\cosh^2\left(\frac{\kappa}{2}\right)}}
}. 
\end{eqnarray*}
The circular-linear correlation, 
which was proposed by \cite{M76} and \cite{JW77}, 
can then be obtained via the formula
\begin{eqnarray*}
R_{x \theta}^2 
&=& 
\frac{r_{xc}^2+r_{xs}^2-2 r_{cs} r_{xc}r_{xs}}{1-r_{cs}^2}.
\end{eqnarray*}

\begin{figure}%[tbh]
\begin{center}
\begin{tabular}{ccc}
\vspace{0.05in} 
(a) $\lambda = 0$ & (b) $\lambda = 0.5$ & (c) $\lambda = 1$ 
\\
\includegraphics[width=5cm,height=5cm]{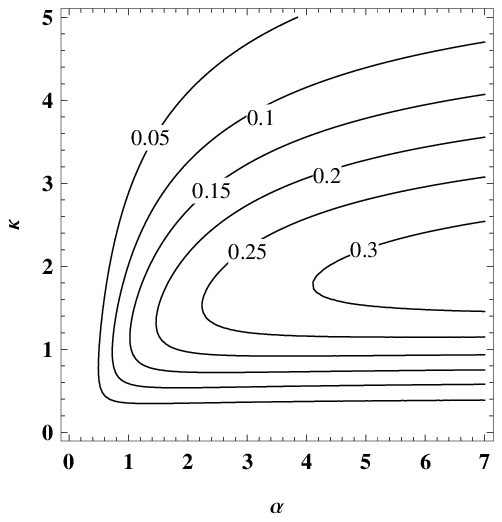}
& 
\includegraphics[width=5cm,height=5cm]{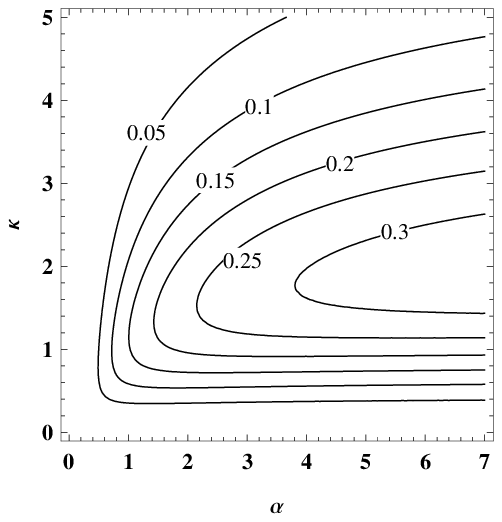}
& 
\includegraphics[width=5cm,height=5cm]{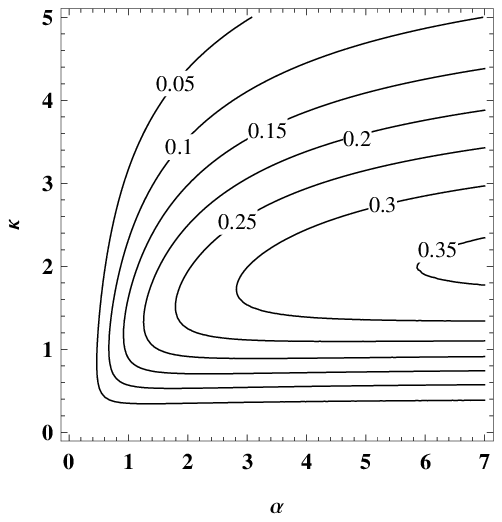}
\end{tabular}
\caption{Contour plots of the circular-linear correlation $R_{x \theta}^2$ as a function of $(\alpha,\kappa)$
over $[0, 7] \times [0, 5]$ 
for 
(a) $\lambda = 0$, (b) $\lambda = 0.5$, (c) $\lambda = 1$. 
}
\label{fig:correlation}
\end{center}
\end{figure}

\begin{figure}%[tbh]
\begin{center}
\begin{tabular}{ccc}
\vspace{0.05in} 
\includegraphics[width=7.5cm,height=5cm]{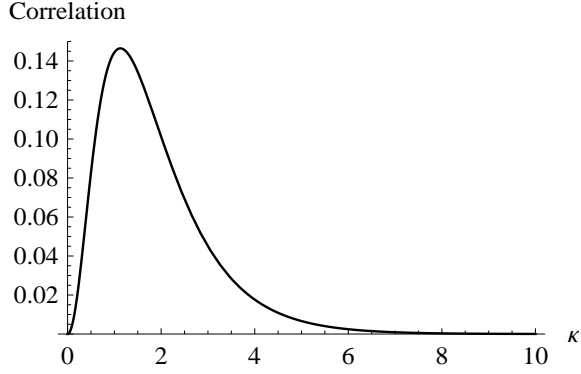}
\end{tabular}
\caption{Circular-linear correlation $R_{x \theta}^2$ for the Johnson-Wehrly model ($\alpha = 1$ and $\lambda = 0$) 
as a function of $\kappa$. 
}
\label{fig:corrJW}
\end{center}
\end{figure}

We see from the above expressions that the circular-linear correlation neither depends on the parameter $\beta$ nor on the sign of $\lambda$. In particular, when $\lambda = 0$, it simplifies to $R_{x \theta}^2 = r_{xc}^2$. The influence of $\alpha$, $\kappa$ and $|\lambda|$ are shown via contour plots  in Figure \ref{fig:correlation}. We observe that, for fixed $\kappa$ and $\lambda$, the correlation increases with $\alpha$. The influence of $\kappa$ goes as follows: at fixed $\alpha$ and $\lambda$, the correlation increases for small values of $\kappa$ until it reaches its maximum, and then decreases.  This phenomenon is directly inherited from the Johnson-Wehrly construction, as we show  in Figure~\ref{fig:corrJW}.

\subsection{A generalization of the WeiSSVM}\label{GGsec}

A natural generalization of our WeiSSVM model consists in replacing the linear Weibull part with the \emph{Generalized Gamma distribution} \cite{S62}, resulting in {the Generalized Gamma sine-skewed von Mises} (GGSSVM) density
\begin{equation}\label{GGSSVM}
(\theta,x)\mapsto  C \left(1+\lambda \sin(\theta-\mu)\right) x^{\alpha-1}\exp\left[-(\beta x)^\gamma\left(1 - \tanh(\kappa)\cos(\theta-\mu)\right)\right], 
\end{equation}
with $\alpha,\gamma,\beta>0$, $\kappa\geq0$, $-1\leq\lambda\leq1$ and $-\pi\leq\mu<\pi$. The normalizing constant is calculated as follows:
\begin{eqnarray*}
&&\int_{-\pi}^{\pi} \int_{0}^{\infty} 
(1+\lambda \sin(\theta-\mu))x^{\alpha-1}\exp[-(\beta x)^\gamma(1 - \tanh(\kappa)\cos(\theta-\mu))] 
dx d\theta\\
&=& 
\frac{\Gamma(\alpha/\gamma)}{\gamma\beta^{\alpha}}
\int_{-\pi}^{\pi} 
\frac{1}{(1-\tanh(\kappa)\cos(\Theta))^{\alpha/\gamma}}
d\theta
\\ 
&=& 
\frac{2\pi\Gamma(\alpha/\gamma)(\cosh(\kappa))^{\alpha/\gamma}P_{\alpha/\gamma-1}(\cosh(\kappa))}{\gamma\beta^{\alpha}}.
\end{eqnarray*}
The WeiSSVM clearly corresponds to $\gamma=\alpha$ in~\eqref{GGSSVM}.  All properties of the GGSSVM   are  obtained along the same lines as our developments in the previous sections, albeit with more involved calculations. It is to be noted that the circular marginal distribution for the GGSSVM is the {sine-skewed Jones--Pewsey} distribution (see \cite{AP11, JP05}). 

We prefer the WeiSSVM over the GGSSVM because of its simplicity, parameter parsimony, higher tractability and its neat link with the Johnson-Wehrly distribution. This explains why the main focus of the present paper lies on the WeiSSVM and why we only briefly mention the density~\eqref{GGSSVM}.

\section{Statistical inference}\label{sec:PE}

\subsection{Parameter estimation}\label{MLE}

Let $( \theta_1,x_1), \ldots, ( \theta_n,x_n)$ be a sample of $n$ independent and identically distributed couples of angular and linear observations drawn from the distribution with density \eqref{WeiSSVM}. 
Then the log-likelihood function can be expressed as

\begin{eqnarray}\label{eq:llf}
&& 
\ell(\alpha, \beta, \mu, \kappa, \lambda) 
= 
(\alpha-1)\sum_{i=1}^{n}\log x_i 
- 
\beta^\alpha\sum_{i=1}^{n}x_i^\alpha\left(1 - \tanh\left(\kappa\right)
\cos(\theta_i-\mu)\right)
\nonumber
\\ 
&& 
+
\sum_{i=1}^{n}\log(1+\lambda \sin(\theta_i-\mu))
+ 
n(\alpha\log\beta+\log\alpha-\log(2\pi\cosh(\kappa))). 
\end{eqnarray}

The elements of the score vector are just the first-order partial derivatives
of (\ref{eq:llf}) with respect to each of the parameters: 
\begin{eqnarray*}
&& 
\frac{\partial \ell}{\partial \alpha}
= 
\sum_{i=1}^{n}\log x_i - \beta^\alpha\sum_{i=1}^{n}\log(\beta x_i)x_i^\alpha\left(1 - \tanh\left(\kappa\right)
\cos(\theta_i-\mu)\right)
+ 
n\left(\log\beta+\frac{1}{\alpha}\right),
\\ 
&& 
\frac{\partial \ell}{\partial \beta}
= 
-\alpha\beta^{\alpha-1}\sum_{i=1}^{n}x_i^\alpha (1 - \tanh(\kappa)\cos(\theta_i-\mu))+\frac{n\alpha}{\beta}, 
\\ 
&& 
\frac{\partial \ell}{\partial \mu}
= 
\beta^\alpha\tanh(\kappa)\sum_{i=1}^{n}x_i^\alpha \sin(\theta_i-\mu)
-\lambda\sum_{i=1}^{n}\frac{\cos(\theta_i-\mu)}{1+\lambda \sin(\theta_i-\mu)}, 
\\ 
&& 
\frac{\partial \ell}{\partial \kappa}
= 
\frac{\beta^\alpha}{(\cosh(\kappa))^2}\sum_{i=1}^{n}x_i^\alpha\cos(\theta_i-\mu)
- 
n\tanh(\kappa), 
\\ 
&&
\frac{\partial \ell}{\partial \lambda}
= 
\sum_{i=1}^{n}\frac{\sin(\theta_i-\mu)}{1+\lambda \sin(\theta_i-\mu)}. 
\end{eqnarray*}
It is difficult to give closed-form expressions for the maximum likelihood estimates (MLEs), hence numerical methods should be used to find the solutions. We used the function {\bf NMaximize} in  {\it Mathematica} where the default numerical maximization algorithm is \emph{Nelder-Mead}, and encountered no problems in the optimization procedure.

\subsection{Submodel and independence testing}\label{sec:tests}

Testing for submodels of the {WeiSSVM model} is straightforward via likelihood ratio tests. For each  parameter $\eta\in\{\alpha,\beta,\mu,\kappa,\lambda\}$, we denote $\hat{\eta}$ the unconstrained maximum likelihood estimate  and $\hat{\eta}_0$ the maximum likelihood estimate under the respective null hypotheses. Two particular instances are of interest. On the one hand, testing for the Johnson-Wehrly submodel, which is taken care of by the test statistic
$$
T_{\rm JW}=  -2(\log\ell(1,\hat\beta_0,\hat\mu_0,\hat\kappa_0,0)-\log\ell(\hat\alpha,\hat\beta,\hat\mu,\hat\kappa,\hat\lambda)),
$$
rejecting $\mathcal{H}_0:(\alpha=1)\cap(\lambda=0)$ at asymptotic level $\eta$ whenever $T_{\rm JW}$ exceeds $\chi^2_{2;1-\eta}$, the $\eta$-upper quantile of the chi-square distribution with 2 degrees of freedom. On the other hand, we are interested in testing for circular-linear independence via the test statistic
$$
T_{\rm Indep}=  -2(\log\ell(\hat\alpha_0,\hat\beta_0,\hat\mu_0,0,\hat\lambda_0)-\log\ell(\hat\alpha,\hat\beta,\hat\mu,\hat\kappa,\hat\lambda)),
$$
to be compared with $\chi^2_{1;1-\eta}$.  Such tests have a long-standing history in the statistical literature; see \cite{GBCGP14} for a recent proposal, based on directional-linear kernel density estimation, and for references.

\subsection{Circular-linear and linear-circular regression}
Since the conditional distributions  take a very simple form, the WeiSSVM model lends itself for circular-linear as well as linear-circular regression, similarly as in~\cite{JW78}. The mean and variance of  $X$ given $\Theta=\theta$ 
correspond to 
$$
{\rm E}[X|\Theta=\theta] = \frac{1}{\beta \left(1 - \tanh(\kappa)\cos(\theta-\mu)\right)^{1/\alpha}}\Gamma\left(\frac{1}{\alpha}+1\right) 
$$
and 
$$
{\rm Var}[X|\Theta=\theta] = 
\frac{1}{\beta^2 \left(1 - \tanh(\kappa)\cos(\theta-\mu)\right)^{2/\alpha}}
\left(
\Gamma\left(\frac{2}{\alpha}+1\right) 
- 
\Gamma\left(\frac{1}{\alpha}+1\right)^2 
\right), 
$$
respectively. 
The first mean direction and mean resultant length of $\Theta$ given $X=x$ are provided by 
$$
\mu_{1|X=x} = \arg\left((\beta x)^\alpha\tanh(\kappa) + i\lambda\right), 
$$
with $\arg$ denoting the argument of a complex number, and \vspace{2mm}
$$
\rho_{1|X=x} = \frac{I_1((\beta x)^\alpha\tanh(\kappa))}{(\beta x)^\alpha\tanh(\kappa)I_0((\beta x)^\alpha\tanh(\kappa))}\sqrt{(\beta x)^{2\alpha}\tanh^2(\kappa)+\lambda^2}, 
$$
respectively. The parameters in each regression model are readily estimated via maximum likelihood, see Section~\ref{MLE}.

\section{Fitting two circular-linear real data sets}\label{sec:Example}

In this section we shall illustrate the good fitting behavior of the WeiSSVM by analyzing two popular data sets from the literature. While the first data set reflects exactly the characteristics of data the WeiSSVM is tailor-made for, namely concentration increasing with length and circular skewness, these attributes are much less marked in the second data set. This allows for a meaningful assessment of the modeling capacities of our new model. In each case, we will compare the WeiSSVM with the Johnson-Wehrly distribution, the independence model, and the GGSSVM of Section~\ref{GGsec}. Our means of comparison shall be the Akaike Information Criterion (AIC) 
and Bayesian Information Criterion (BIC), and  we will apply the tests of Section~\ref{sec:tests}.

Besides these alternative distributions, we shall as well indicate a comparison with the Mardia-Sutton model \cite{MS78} and its recent extension proposed in~\cite{KS08}, the Kato-Shimizu distribution.
The latter  has as density 
\begin{equation}\label{pdf:KS} 
f_{KS}(\theta, x) 
= 
C \exp\left[
-\frac{(x-\mu(\theta))^2}{2\sigma^2} 
+ \kappa_1 \cos(\theta-\mu_1)
+ \kappa_2 \cos(2(\theta-\mu_2))
\right], 
\end{equation}
where $-\pi\le\theta<\pi$, $-\infty<x<\infty$, 
$\sigma>0$, $\kappa_1, \kappa_2 >0$, $-\pi\le \mu_1 <\pi$, $-\pi/2\le \mu_2 <\pi/2$, 
$\mu(\theta)=\mu'+\lambda\cos(\theta-\nu)$, $-\infty<\mu'<\infty$, 
$\lambda>0$, $-\pi\le\nu<\pi$ and its normalizing constant $C$ is provided by 
$$
C^{-1} 
= 
(2\pi)^{3/2}\sigma\left(
I_0(\kappa_1)I_0(\kappa_2) 
+ 
2 \sum^{\infty}_{j=1}I_j(\kappa_2)I_{2j}(\kappa_1)\cos(2j(\mu_1-\mu_2)) 
\right). 
$$
The Mardia-Sutton model is obtained by setting $\kappa_2=0$ in~\eqref{pdf:KS}. The infinite sum in the normalizing constant then vanishes, resulting in a simpler density. In the following comparisons, we draw the reader's attention to the fact that  both the Mardia-Sutton and Kato-Shimizu models are defined over $\R\times\mathcal{C}_1$ with $\mathcal{C}_1$ the unit circle in $\R^2$, whereas the above-mentioned densities are  defined over $\R^+\times\mathcal{C}_1$.

%We consider  two well-known data sets that have already been investigated in numerous preceding papers. 
%The third example is new, and hence we will study it in more details.

\subsection{Periwinkle data}

%As first illustrative example, 
We give an analysis of   $n = 31$ observations which consist of the movements of blue periwinkles after they had been transplanted downshore from the height at which they normally live. The data set was taken from Table 1 of \cite{FL92}; see that paper for details about the experience.

\begin{figure}[tbh]
\begin{center}
\includegraphics[width=8cm,height=8cm]{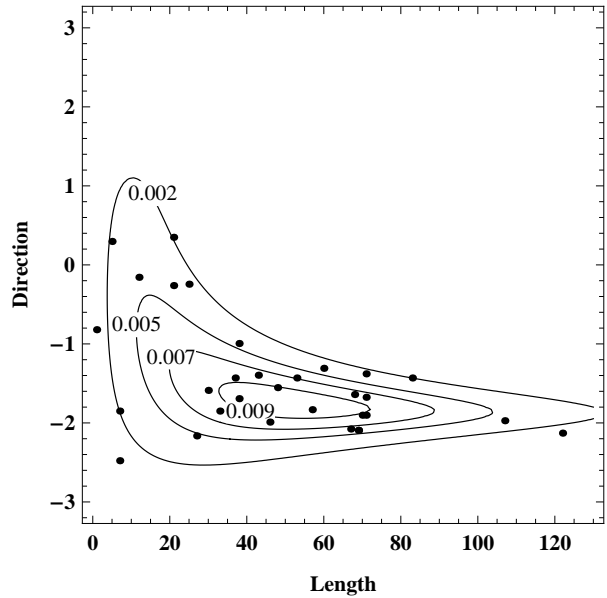}
\caption{Contour plot of the blue periwinkle data (in lengths and radians), 
together with the fitted WeiSSVM density.
The data are plotted over $[0, 125)\times[-\pi, \pi) $, with the distance indicated in cm. } 
\label{peri:fitted}
\end{center}
\end{figure}

A visual inspection of the data points in Figure~\ref{peri:fitted}  reveals that the  concentration of the circular part tends to increase with length, which is precisely one of the features that the WeiSSVM model can well incorporate. Moreover, \cite{KS08} have shown that, on basis of the Pewsey test of symmetry (see \cite{P02}), the circular part of the data is asymmetric.

Table \ref{table:periwinkle} presents 
the maximum likelihood estimates, maximized log-likelihood, Akaike and Bayesian Information Criterion values obtained from all models under investigation. 
As we can see, the location parameters of the GGSSVM and its submodels are close 
(note that the location of the Independence model is $-2.97+\pi/2 = -1.40$, as explained at the end of Section~\ref{model}) and 
the WeiSSVM has the lowest AIC and BIC values. It clearly improves on Johnson-Wehrly and Mardia-Sutton, and even on the flexible Kato-Shimizu model. It is quite remarkable to notice the tiny difference in the maximized log-likelihood between WeiSSVM and the embedding model, the GGSSVM. 
The likelihood ratio test for the Johnson-Wehrly submodel (w.r.t. the WeiSSVM) takes value $T_{\rm JW}=-2(-182.93+168.57)=28.72$, with $p$-value $=0.00$, which emphatically rejects the Johnson-Wehrly model. Even stronger, the independence test yields $T_{\rm Indep}=-2(-187.25+168.57)=37.36$,  stressing the dependence between the angular and the linear part.

\begin{table}[tbp]
\caption{Maximum likelihood estimates, maximized
log-likelihood (MLL), Akaike Information Criterion (AIC) 
and Bayesian Information Criterion (BIC) values 
for the Weibull sine-skewed von Mises (WeiSSVM) 
and its competitor models, the Generalized Gamma sine-skewed von Mises (GGSSVM),  the Johnson-Wehrly (JW), the independence (Indep.), Mardia-Sutton (MS) and Kato-Shimizu (KS) models, fitted to the blue periwinkle data. 
}\label{table:periwinkle}
\begin{center}
\small
\begin{tabular}{lrrlrrrllrrr}
Distributions & \multicolumn{1}{l}{$\hat{\alpha}$} & \multicolumn{1}{l}{$\hat{\beta}$} & $\hat{\gamma}$ & \multicolumn{1}{l}{$\hat{\mu}$} & \multicolumn{1}{l}{$\hat{\kappa}$} & \multicolumn{1}{l}{$\hat{\lambda}$} &  &  & \multicolumn{1}{l}{MLL} & \multicolumn{1}{l}{AIC} & \multicolumn{1}{l}{BIC} \\ \hline
WeiSSVM & \phantom{0}2.01 & \phantom{0}0.05 &  & \llap{$-$}1.90 & 1.68 & 1.00 &  &  & \llap{$-$}168.57 & 347.13 & 354.30 \\
GGSSVM & \phantom{0}2.00 & \phantom{0}0.05 & \multicolumn{1}{r}{\phantom{0}2.04} & \llap{$-$}1.90 & 1.70 & 1.00 &  &  & \llap{$-$}168.56 & 349.13 & 357.73 \\
JW &  & \phantom{0}0.10 &  & \llap{$-$}1.70 & 1.44 &  &  &  & \llap{$-$}182.93 & 371.86 & 376.16 \\
Indep. & \phantom{0}1.54 & \phantom{0}0.02 &  & \llap{$-$}2.97 &  & 1.00 &  &  & \llap{$-$}187.25 & 382.50 & 388.24 \\
 &  &  &  &  &  &  &  &  &  &  &  \\
 & \multicolumn{1}{l}{$\hat{\mu}$} & \multicolumn{1}{l}{$\hat{\sigma}$} & $\hat{\lambda}$ & \multicolumn{1}{l}{$\hat{\nu}$} & \multicolumn{1}{l}{$\hat{\mu}_1$} & \multicolumn{1}{l}{$\hat{\mu}_2$} & $\hat{\kappa}_1$ & $\hat{\kappa}_2$ & \multicolumn{1}{l}{MLL} & \multicolumn{1}{l}{AIC} & \multicolumn{1}{l}{BIC} \\ \cline{2-12}
MS & 28.58 & 24.43 & \multicolumn{1}{r}{29.63} & \llap{$-$}2.11 & \llap{$-$}1.52 &  & \multicolumn{1}{r}{2.59} &  & \llap{$-$}176.88 & 365.75 & 374.36 \\
KS & 28.58 & 24.43 & \multicolumn{1}{r}{29.63} & \llap{$-$}2.11 & \llap{$-$}0.97 & 0.73 & \multicolumn{1}{r}{8.16} & \multicolumn{1}{r}{3.46} & \llap{$-$}168.46 & 352.93 & 364.40
\end{tabular}
\end{center}
\end{table} %

As a conclusion, our WeiSSVM model (with $5$ parameters) is a good-fitting and parsimonious model for the periwinkle data set. For visual impression, we have superimposed the contour plot of the fitted WeiSSVM model on a {scatter} plot of the data in the panel making up Figure \ref{peri:fitted}. 

\subsection{Wind direction and temperature data}
\begin{figure}[tbh]
\begin{center}
\includegraphics[width=8cm,height=8cm]{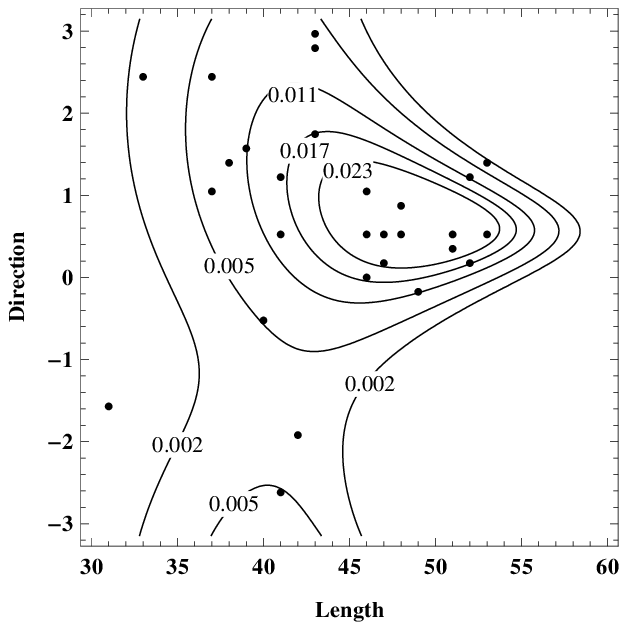}
\caption{Contour plot of the wind and temperature data (in lengths and radians), 
together with the fitted WeiSSVM density.
The data are plotted over $[30, 60)\times[-\pi, \pi)$, with temperature indicated in Fahrenheit. } 
\label{ozone:fitted}
\end{center}
\end{figure}

As  second example, we consider the original data set from \cite{MS78}, consisting of 28 measurements of wind direction and temperature at Kew during the period 1956-1960. The data are taken from Table 1 in \cite{MS78} and illustrated in Figure~\ref{ozone:fitted}. Although the effect noticed for the periwinkle data, namely high concentration for high linear values, is less marked here, Mardia and Sutton have noted (and established) a strong dependence between the circular and the linear component.

It has been shown in \cite{KS08} that the Mardia-Sutton model is extremely good for this data set; it is therefore very interesting to compare it with the WeiSSVM. Table~\ref{table:MSwind} contains the 
maximum likelihood estimates, maximized log-likelihood, Akaike and Bayesian information criterion values. 
The (circular) location parameters of our proposed models are almost the same  
(again, the location of the Independence model is $-0.62 + \pi/2 = 0.95$).  
We see that our WeiSSVM model best incorporates the non-trivial behavior of this data set (its AIC and BIC values are clearly below that of the MS model), 
and again it is much better than the Johnson-Wehrly model (which is clearly rejected as submodel).  A contour plot of the fitted WeiSSVM model with a {scatter} plot of the data is provided in Figure~\ref{ozone:fitted}. 
We finally note that the independence test of course heavily rejects ($p$-value $=0.00$) the null of independence, hereby agreeing with~\cite{MS78}.

\begin{table}[tbp]
\caption{Maximum likelihood estimates, maximized
log-likelihood (MLL), Akaike Information Criterion (AIC) 
and Bayesian Information Criterion (BIC) values 
for the Weibull sine-skewed von Mises (WeiSSVM) 
and its competitor models, the Generalized Gamma sine-skewed von Mises (GGSSVM),  the Johnson-Wehrly (JW), the independence (Indep.), Mardia-Sutton (MS) and Kato-Shimizu (KS) models, fitted to the wind-temperature data. 
}\label{table:MSwind}
\begin{center}
\small
\begin{tabular}{lrrlrrrllrrr}
Distributions & \multicolumn{1}{l}{$\hat{\alpha}$} & \multicolumn{1}{l}{$\hat{\beta}$} & $\hat{\gamma}$ & \multicolumn{1}{l}{$\hat{\mu}$} & \multicolumn{1}{l}{$\hat{\kappa}$} & \multicolumn{1}{l}{$\hat{\lambda}$} &  &  & \multicolumn{1}{l}{MLL} & \multicolumn{1}{l}{AIC} & \multicolumn{1}{l}{BIC} \\ \hline
WeiSSVM & 10.72 & 0.02 &  & 0.54 & 1.10 & 0.49 &  &  & \llap{$-$}125.70 & 261.39 & 268.05 \\
GGSSVM & 10.78 & 0.02 & \multicolumn{1}{r}{10.64} & 0.54 & 1.09 & 0.49 &  &  & \llap{$-$}125.69 & 263.39 & 271.38 \\
JW &  & 0.03 &  & 0.83 & 0.60 &  &  &  & \llap{$-$}180.77 & 367.55 & 371.54 \\
Indep. & \phantom{0}8.90 & 0.02 &  & \llap{$-$}0.62 &  & 0.78 &  &  & \llap{$-$}134.32 & 276.64 & 281.97 \\
 &  &  &  &  &  &  &  &  &  &  &  \\
 & \multicolumn{1}{l}{$\hat{\mu}$} & \multicolumn{1}{l}{$\hat{\sigma}$} & $\hat{\lambda}$ & \multicolumn{1}{l}{$\hat{\nu}$} & \multicolumn{1}{l}{$\hat{\mu}_1$} & \multicolumn{1}{l}{$\hat{\mu}_2$} & $\hat{\kappa}_1$ & $\hat{\kappa}_2$ & \multicolumn{1}{l}{MLL} & \multicolumn{1}{l}{AIC} & \multicolumn{1}{l}{BIC} \\ \cline{2-12}
MS & 42.07 & 4.86 & \multicolumn{1}{r}{\phantom{0}5.01} & 0.36 & 0.88 &  & \multicolumn{1}{r}{1.14} &  & \llap{$-$}128.10 & 268.19 & 276.19 \\
KS & 42.07 & 4.86 & \multicolumn{1}{r}{\phantom{0}5.01} & 0.36 & 1.09 & 0.48 & \multicolumn{1}{r}{1.02} & \multicolumn{1}{r}{0.53} & \llap{$-$}126.66 & 269.32 & 279.98
\end{tabular}
\end{center}
\end{table} %

\section{Future research and a directional-linear extension}\label{final}

Given its very good fitting capacities and simple parameter interpretation, the WeiSSVM is a viable model to investigate in detail further data sets. Two concrete examples shall be elucidated in the future. The first concerns ecological data related to trees. Indeed, \cite{AKSAK12} have only used the direction of fallen logs, hence a pure circular setting, to model the influence of neighborhood structure and directionality of radiation on crown asymmetry; a more detailed analysis can be obtained by adding as linear part the distance to each neighboring tree. 
%\cite{ASAKK15} have studied the tree crown vectors by using one of the models from \cite{JW78}; this analysis shall be improved by means of the WeiSSVM. 
The second data set concerns  cylindrical data consisting of the burnt area and the direction of wildfires in Portugal, as analyzed in \cite{BPL12} and \cite{GBCGP14}. Our parametric model will be an interesting alternative especially to the non-parametric approach of the latter paper. 

The latter data are both circular-linear and directional-linear, hence requiring an extension of the WeiSSVM to the directional-linear setting. We shall now briefly indicate how straightforward this extension is. It   is obtained by replacing the circular sine-skewed von Mises density with its equivalent on unit spheres 
$\mathcal{S}^{k-1}=\{\vb\in\R^k:\|\vb\|=1\}$, $k\geq 3$, 
recently defined in \cite{LV14}. The cosine part simply becomes the scalar product $\thetab'\mub$ between $\thetab\in\mathcal{S}^{k-1}$ and the location parameter $\mub\in\mathcal{S}^{k-1}$, while $\lambda\sin(\theta-\mu)$  is expressed as $\sqrt{1-(\thetab'\mub)^2}\lambdab'\Sb_{\mub}(\thetab)$, $\lambdab\in\mathcal{S}^{k-1}$, where $\Sb_{\mub}=(\thetab-(\thetab'\mub)\mub)/\|\thetab-(\thetab'\mub)\mub\|$ is the multivariate sign vector on the unit sphere. We refer to \cite{LV14} for further information. 

%\subsection{The WeiSSFVML and its properties}

The density of the Weibull sine-skewed Fisher-von Mises-Langevin\footnote{In higher dimensions, the VM is called  Fisher-von Mises-Langevin and hence abbreviated FVML.},  in short WeiSSFVML, distribution on $\mathcal{S}^{k-1}\times\R^+$, for the directional part with respect to the usual surface area measure $d\sigma_{k-1}$, is defined as
\begin{eqnarray}\label{WeiSSVMF}
(\thetab,x)\mapsto f(\thetab,x)=C_k \left(1+\sqrt{1-(\thetab'\mub)^2}\lambdab'\Sb_{\mub}(\thetab)\right)x^{\alpha-1} \exp\left[-(\beta x)^\alpha(1-\tanh(\kappa)\thetab'\mub)\right].
\end{eqnarray}
The normalizing constant of the distribution (\ref{WeiSSVMF}) is simply given by 
$$
C_k
= 
\frac{\alpha\beta^\alpha(\sinh(\kappa))^{(k/2)-1}}
{(2\pi)^{k/2}\cosh(\kappa)P_{k/2-2}^{1-(k/2)}(\cosh(\kappa))}. 
$$
Indeed 
\begin{eqnarray*}
&&
\int_{S^{k-1}} \int_{0}^{\infty} 
\left(1+\sqrt{1-(\thetab'\mub)^2}{\lambdab}'{\Sb}_{\mub}(\thetab)\right)
x^{\alpha-1}\exp\left[{-(\beta x)^\alpha(1 - \tanh(\kappa)\thetab'\mub)}\right]
dx d\sigma_{k-1}(\thetab)
\\
&=&
\int_{S^{k-1}} \int_{0}^{\infty} x^{\alpha-1}\exp\left[-(\beta x)^\alpha(1 - \tanh(\kappa)\thetab'\mub)\right]dx d\sigma_{k-1}(\thetab)
\\
&=&\int_{S^{k-1}(\mub^\perp)}\int_{-1}^1 \int_{0}^{\infty} x^{\alpha-1}\exp\left[{-(\beta x)^\alpha(1 - \tanh(\kappa)t)}\right]dx d\sigma_{k-2}({\vb}) (1-t^2)^{(k-3)/2}dt
\\
&=& 
\frac{2\pi^{k/2}}{\alpha\beta^\alpha \Gamma(k/2)B(1/2,(k-1)/2)}
\int^{1}_{-1} 
\frac{(1-t^2)^{(k-3)/2}}{1-\tanh(\kappa)t}
dt
\\ 
&=& 
\frac{(2\pi)^{k/2}\cosh(\kappa)P_{k/2-2}^{1-(k/2)}(\cosh(\kappa))}
{\alpha\beta^\alpha(\sinh(\kappa))^{(k/2)-1}}, 
\end{eqnarray*}
where $B(\cdot, \cdot)$ denotes the beta function. We have used above the change of variables formula 
$d\sigma_{k-1}(\thetab)=(1-t^2)^{(k-3)/2}d\sigma_{k-2}({\vb})dt$ 
where 
$\vb\in\mathcal{S}^{k-1}(\mub^\perp)=\{\vb\in\R^k:\|{\vb}\|=1,{\vb}'{\mub} = 0\}$, 
the equality $\omega_{k-1}=\omega_{k}/B(1/2,(k-1)/2)$ (with $\omega_k=2\pi^{k/2}/\Gamma(k/2)$ the surface area measure of $\mathcal{S}^{k-1}$) as well as, like for  the result of Section~5 in \cite{JP05},  the following relationship of the associated Legendre function 
(equation 8.711.1 of \cite{GR15}, p. 969)
$$
P^{-\mu}_{\nu}(z) 
=
\frac{(z^2-1)^{\frac{\mu}{2}}}{2^\mu\sqrt{\pi}\Gamma(\mu+\frac{1}{2})} 
\int^{1}_{-1} 
\frac{(1-t^2)^{\mu-\frac{1}{2}}}{(z+t\sqrt{z^2-1})^{\mu-\nu}}
dt
\quad 
[\Re \mu > -1/2, \quad |\arg(z\pm 1)|<\pi]. 
$$

Clearly, the distribution (\ref{WeiSSVMF}) reduces to (\ref{WeiSSVM}) when $k=2$, 
and as in \cite{JP05}, 
the distribution also has a simpler form when $k=3$, namely, 
\begin{eqnarray*}
f(\thetab, x)=\frac{\alpha\beta^\alpha\tanh(\kappa)}{4\pi\kappa} \left(1+\sqrt{1-(\thetab'\mub)^2}\lambdab'\Sb_{\mub}(\thetab)\right) x^{\alpha-1} \exp\left[-(\beta x)^\alpha(1-\tanh(\kappa)\thetab'\mub)\right].
\end{eqnarray*}
Investigating stochastic properties as well as estimation procedures is beyond the scope of the present paper and left for future research.

\section*{Acknowledgements}
Toshihiro Abe was supported in part by JSPS KAKENHI Grant Number 15K17593 
and Nanzan University of Pache Research Subsidy I-A-2 
for the 2015 academic year. 
Christophe Ley was supported in part by the Fonds National de la Recherche Scientifique, Communaut\'e fran\c caise de Belgique,  via a Mandat de Charg\'e de Recherche.

\bibliographystyle{abbrv}
\bibliography{biblio_AL}

%/Users/christopheley/Desktop/Tout/Articles/Papers finished/WeiSSVM3paper/
%%%  end of document  %%%
\end{document}